\documentclass[showpacs,preprintnumbers,amsmath,amssymb,twocolumn,10pt, superscriptaddress]{revtex4-1}
\usepackage{color}
\usepackage{amsmath,amssymb}
\usepackage{dsfont}
\usepackage{pgfplots}
\usepackage{pgf,tikz}
\usetikzlibrary{arrows,snakes,backgrounds}
\usepackage{epsfig,graphicx,overpic}

\begin{document}
\title{Atomic quantum memory for multimode frequency combs}

\author{Zhan Zheng}
\affiliation{Laboratoire Kastler Brossel, Sorbonne Universit\'e-UPMC, ENS, Coll\`ege de France, CNRS, 4 place Jussieu, 75252 Paris, France}
\affiliation{Key Laboratory for Precision Spectroscopy, East China Normal University, Shanghai 200062, China}
\author{Oxana Mishina}
\affiliation{Theoretische Physik, Universit\"at des Saarlandes, D-66041 Saarbr\"ucken, Germany}
\author{Nicolas Treps}
\affiliation{Laboratoire Kastler Brossel, Sorbonne Universit\'e-UPMC, ENS, Coll\`ege de France, CNRS, 4 place Jussieu, 75252 Paris, France}
\author{Claude Fabre}
\affiliation{Laboratoire Kastler Brossel, Sorbonne Universit\'e-UPMC, ENS, Coll\`ege de France, CNRS, 4 place Jussieu, 75252 Paris, France}

\begin{abstract}
We propose a Raman quantum memory scheme that uses several atomic ensembles to store and retrieve the multimode highly entangled state of an optical quantum frequency comb, such as the one produced by parametric down-conversion of a pump frequency comb. We analyse the efficiency and the fidelity of such a quantum memory. Results show that our proposal may be helpful to multimode information processing using the different frequency bands of an optical frequency comb. \vspace{5mm}
\end{abstract}


\maketitle

Quantum information is a fascinating subject, as it makes use of the deepest aspects of quantum theory, which is inherently an information theory. When the information is carried by quantum states belonging to a high-dimensional Hilbert space, for example by quantum states of highly multimode light, one expects a potential significant increase in the information capacity. In the toolbox of quantum information processing, one of the most important tools is the quantum memory, without which all the advantages of quantum information cannot be fully exploited. Quantum memories are actively studied, both at the theoretical and experimental level \cite{QM1}, because they constitute an important resource in long-distance quantum communication. 

Up to now these studies concerned mainly quantum states contained in a single pulse of single transverse mode light, i.e., single-mode configurations. They also have been extended to some kinds of multimode fields, either in the spatial or temporal domain. In the spatial domain, the quantum aspects of highly multimode light have been actively studied in the past years \cite{Kolobov}. In particular, spatially multimode quantum memories have been designed and built \cite{QM2}. In the temporal domain, multimode quantum memories have been developed to store long pulses of different mean frequencies \cite{QM3,Afzelius,Usmani,Reim2010}. Another possibility is to store different modes consisting of different time slots \cite{Bonarota} or different temporal shapes of short light pulses \cite{Golubeva2011}, for which efficient pulse-shaping techniques exist. This is the reason why we have chosen to explore the problem of storing quantum states of "optical frequency combs": Highly multimode quantum frequency combs  have indeed been recently experimentally implemented \cite{Roslund2014}, and have been shown to exhibit genuine multipartite \cite{Pinel,Medeiros} and full entanglement \cite{Gerke}. Such highly multimode quantum states are a promising resource for quantum information processing and measurement-based quantum computing \cite{Hayes,Medeiros,Ferrini,Ferrini2,Campbell}.

\begin{figure}
\centering
\includegraphics[scale=.4]{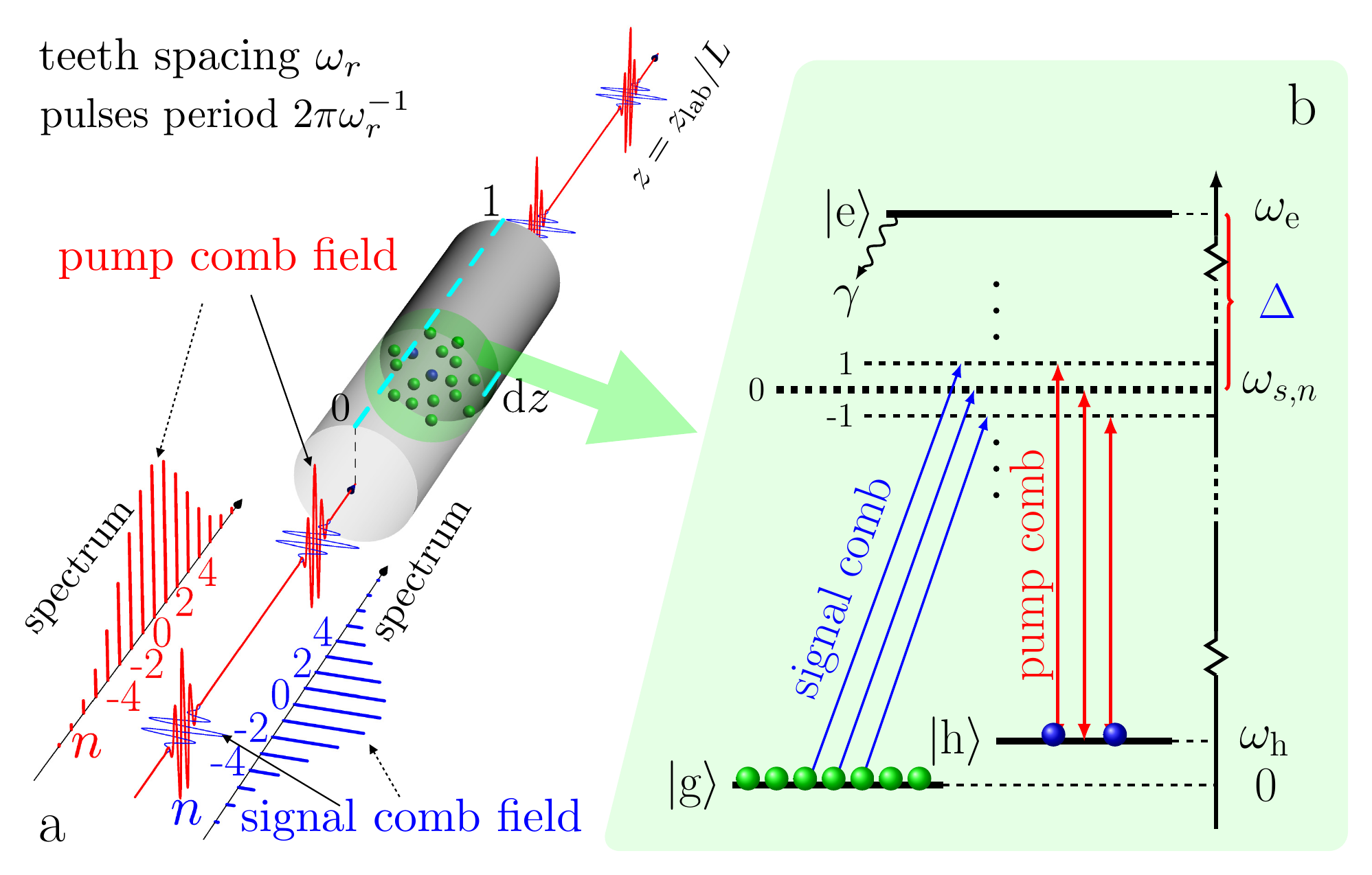}
\caption{\footnotesize (Color online) a. Short pulses interacting with an atomic ensemble; b. Three-level $\Lambda$-type atoms coupled to comb fields via perfect two-photon resonant channels in an off-resonant Raman configuration.}\label{fig1}
\end{figure}

An ideal optical frequency comb consists of a series of periodic phase-coherent short pulses, which correspond in the frequency domain to a series of equally spaced phase-locked monochromatic components, hence the name of "combs" [Fig.\ref{fig1}(a)]. Typically, the frequency difference between neighboring teeth $\omega_r/2\pi$ is on the order of 10 MHz to 1 GHz, whereas the teeth extend over a spectral domain of 10 nm for 100 fs pulses in the visible, and 0.1 nm for 10 ps pulses. Let us note that the number of teeth, and therefore of spectral modes, is very large, between $10^3$ and $10^6$, so that this kind of light is likely to carry a very large number of frequency modes. Note that the memory for optical frequency combs that we consider here must not be confused with the atomic frequency comb memory \cite{Afzelius}.

The aim of our Rapid Communication is to see whether the multipartite entanglement between the frequency modes can be transferred to an atomic ensemble. This is nontrivial since (1) the bandwidth associated with the frequency combs far exceeds the typical bandwidth of atomic systems, and (2) an intrinsic multimode state \cite{Treps2005} may not be stored and retrieved by an intrinsic single-mode drive. Nevertheless, we show here that by choosing a suitable set of classical drives it is still possible to map the phase correlations of the optical frequency comb into the atomic memories, and to retrieve them in the reading process.

%===========================AtomicEnsemble===============================%
%===========================AtomicEnsemble===============================%

We consider an extended ensemble of $N_a$ three-level atoms in the $\Lambda$ configuration with dipole $\hat{\bf d}$, uniformly contained in a volume  of length $L$ elongated in the $z_{\text{lab}}$ direction, interacting  in the Raman configuration with two frequency combs, written as $\hat{\textbf{E}}_{\mu}^{(+)}(z,t)=\text{i}\textbf{e}_{\mu}e^{-\text{i}\omega_{\mu}t}\mathcal{E}_{\mu}\hat{E}_{\mu}^{(+)}(z,t)$ with $\mu=s,p,$ and $\textbf{e}_{\mu}, \omega_{\mu}, \mathcal{E}_{\mu}$ the corresponding polarization, central frequency, and electric field unit which depends on the central frequency and the beam cross section, in the frame moving with the pulses $(z=z_{\text{lab}}/L,t=t_{\text{lab}}-z_{\text{lab}}/c)$, that propagate along the $z_{\text{lab}}$ axis (Fig. \ref{fig1}a). The signal comb field operator writes
\begin{equation}
\hat{E}_s^{(+)}(z,t)=\sum_m\hat{s}_{m}(z,t)e^{-\text{i}m\omega_rt},
\end{equation}
where $\hat{s}_{m}(z,t)$ is a slowly varying annihilation operator, obeying 
\begin{equation}
[\hat{s}_m(z,t),\hat{s}_n^{\dag}(z',t')]=\delta_{mn}\delta(t-t').\label{FCCR}
\end{equation}

As is known for long pulses, the signal field can be stored in the spatial dependence of the coherence  between the ground state $|$g$\rangle$ and the metastable state $|\text{h}\rangle$ of an ensemble of atoms via a two-photon Raman transition induced by a classical pump comb field $\hat{E}_p^{(+)}(z,t)=\alpha_p\sum_{m} p_{m}e^{-\text{i}m\omega_rt}$. In this expression $\alpha_p$ is the field amplitude and $\sum_m |p_m|^2=1$. We assume perfect Raman resonance and a large single-photon detuning $\Delta=\omega_e-\omega_s$, where $\omega_e$ is the Bohr frequency of the excited state $|$e$\rangle$ with linewidth $\gamma$ [see Fig. \ref{fig1}(b)].

By destroying a photon in the signal comb field and emitting a stimulated photon in the other, the spectral coherence of the comb will be transferred to the $|\text{g}\rangle\langle \text{h}|$ coherence. To describe this process, let us introduce a collective atomic operator
\begin{equation}
\hat{b}(z,t)=-\text{i}\frac{e^{-\text{i}(\omega_s-\omega_p)t}}{\sqrt{N_a} \text{d}z}\sum_{j=1}^{N_a\text{d}z}|\text{g}\rangle_{j}\langle \text{h}|_j,
\end{equation}
with $j$ labeling all the atoms in a layer of infinitesimal thickness d$z$ at $z$ [see Fig. \ref{fig1}(a)]. The nonvanishing equal-time bosonic commutation relation of collective operators of slices at $z$ and $z'$ obeys
$[\hat{b}(z,t), \hat{b}^{\dag}(z',t)]%=\frac{\delta_K(z,z')}{\rho(\text{d}z)^2}\sum_{j=1}^{\rho\text{d}z}(|\text{g}\rangle_{j}\langle\text{g}|-|\text{h}\rangle_j\langle \text{h}|)\simeq 
\xrightarrow{\text{d}z\rightarrow0}\delta(z-z')$. If the number of photons in the input signal field is much less than the number of atoms, and all atoms are initially in the ground state, the coupled atom-field equations are 
\begin{subequations}
\begin{eqnarray}
& (\partial_{t}+\gamma_s)\hat{b}(z,t)=(d\gamma_s)^{\frac{1}{2}}{\bf p}^{\text{H}}[\hat{\bf s}(z,t)-\hat{\bf f}(z,t)], \label{Eqforb}&\\
&\partial_{z}\hat{\bf s}(z,t)=-(d\gamma_s)^{\frac{1}{2}}{\bf p}\hat{b}(z,t).& \label{EqforField}
\end{eqnarray}
\end{subequations}
These equations are analogous to the ones in Ref. \cite{Kozhekin2000} except that in order to treat the combs, we have used a vectorial approach where the vector column notation $\hat{\bf v}=\text{col}(\cdots,\hat{v}_{-1},\hat{v}_{0},\hat{v}_{1},\cdots)$ describes in a compact way the different "teeth" of the $\hat{\bf v}$ comb, where $\hat{\bf  v}$ stands for the pump, signal, and noise combs ${\bf p}$, $\hat{\bf s}$, and $\hat{\bf f}$. The superscripts $ ^{\text{H}}$ and $ ^{\text{T}}$ of a vector represent the Hermitian conjugate and transpose respectively in this Rapid Communication. Here, $d\equiv N_a\langle\text{e}|\hat{\bf d}\cdot{\bf e}_s|\text{g}\rangle^2 \mathcal{E}_{s}^2/\gamma\hbar^{2}$ is the on-resonance optical depth; $\gamma_s\equiv \gamma|\Delta^{-1}\Omega_p|^2$, with $\Omega_p\equiv\hbar^{-1}\langle\text{e}|\hat{\bf d}\cdot{\bf e}_p|\text{h}\rangle \mathcal{E}_{p}\alpha_p$ being the Rabi frequency of the pump field, is the induced decay rate, which vanishes when the pump field is off. Stark shifts, which come from Rayleigh scatterings and can be compensated \cite{Nunn2007}, have been removed in Eqs. (\ref{Eqforb}) and (\ref{EqforField}). $\hat{\bf f}$ is a vectorial Langevin noise operator inserted to preserve the commutation relations of the field and atoms. At room temperature and in the optical domain, $\langle \hat{f}_{m}\hat{f}_{n}\rangle=\langle \hat{f}^{\dag}_{m}\hat{f}_{n}\rangle =0$ for any $m,n$. 

The full solutions of Eqs. (\ref{Eqforb}) and (\ref{EqforField}), which can be obtained with the Laplace transform technique, are complicated, as in Ref. \cite{Kozhekin2000}. We will here focus our analysis on the multimode uncorrelated squeezed "supermodes" \cite{Valcarcel} that can be identified in the signal frequency comb by diagonalizing the covariance matrix. They are characterized by the eigenvalues of this matrix \cite{Arvind}. The elements in a covariance matrix are an average of quadratic operators over the vacuum state. By rearranging these quadratic operators in normal ordering form, the average of all terms related to the vacuum inputs and to the Langevin noises vanishes \cite{Golubev}. Therefore, it is not necessary to keep the noise-induced terms in the following solutions for the linear dynamics. At time $T$, the end of the storage process, the $z$-dependent coherence operator in the ensemble $\hat{b}(z,T)$ can be shown, after some calculation, to be
\begin{equation}
\hat{b}(z,T)=\sqrt{\alpha\over{T}}\int_{0}^{T}\!\! e^{-\gamma_st}J_{0}(2\sqrt{\frac{\alpha zt}{T}}){\bf p}^{\text{H}}\hat{\bf s}^{\text{in}}(T\!-\!t)\text{d}t,\label{BCoh}
\end{equation}
where  $J_0$ is the zeroth-order Bessel function, $\alpha=d\gamma_sT$, and $\hat{\bf s}^{\text{in}}$ is the field annihilation operators that carry the information of the multimode incident light. Note that the expression (\ref{BCoh}) is the vectorial analogue of the term in Ref. \cite{Kozhekin2000}. Contrary to a single-pulse case \cite{Reim2011}, the actual interaction time during the comb storage is much shorter than the pulse duration $T$, which counter-intuitively but advantageously avoids the constraint for the induced decay rate $\gamma_s$ being smaller than $1/T$.

Let us now turn to the reading process, in which another pump field ${\bf p}_{\text{read}}$ is applied to the atomic medium. The signal field annihilation operator column vector at $z=1$ reads \cite{Kozhekin2000}
\begin{alignat}{2}
\hat{\bf s}^{\text{out}}(t)=&
-{\bf p}_{\text{read}}e^{-\gamma_{s}t}\sqrt{\alpha\over{T}}\int_{0}^{1}\text{d}yJ_{0}(2\sqrt{\alpha\frac{1-y}{T}t})\hat{b}(y,T).\label{QField} &&\notag
\end{alignat}
It indicates that the entanglement in the write-in state can be manipulated and redistributed in other modes. Taking ${\bf p}_{\text{read}}={\bf p}$ (the reading beam identical to the writing beam), the input-output relation in the Fourier frequency domain turns out to be
\begin{subequations}
\begin{eqnarray}
&\underline{\hat{\bf s}}^{\text{out}}(\omega)=-
\mathcal{K}_{\omega}e^{\text{i}\omega T}{\bf p}{\bf p}^{\text{H}}\underline{\hat{\bf s}}^{\text{in}}(\omega),&  \label{in-out}\\
&\mathcal{K}_{\omega}= 1-e^{-\frac{d\gamma_s}{\gamma_{s}+\text{i}\omega}},&
\end{eqnarray}
\end{subequations}
where the underlines indicate Fourier transforms. The conclusion is that one can indeed retrieve at the output a copy of the input field, but with the constraint that the retrieved light is in the single mode ${\bf p}$ because of the rank-1 projector ${\bf pp}^{\text{H}}$. At large optical depth $d$, $\mathcal{K}_{\omega}\rightarrow1$, therefore $\underline{\hat{\bf s}}^{\text{out}}(\omega)\rightarrow-e^{\text{i}\omega T}{\bf p}{\bf p}^{\text{H}}\underline{\hat{\bf s}}^{\text{in}}(\omega)$. This represents the first result of this Rapid Communication: One is able to store and retrieve a squeezed frequency comb of any modal shape by choosing the proper pump field shape.

%===========================AtomicEnsemble===============================%
\begin{figure}
\centering
\includegraphics[scale=.27]{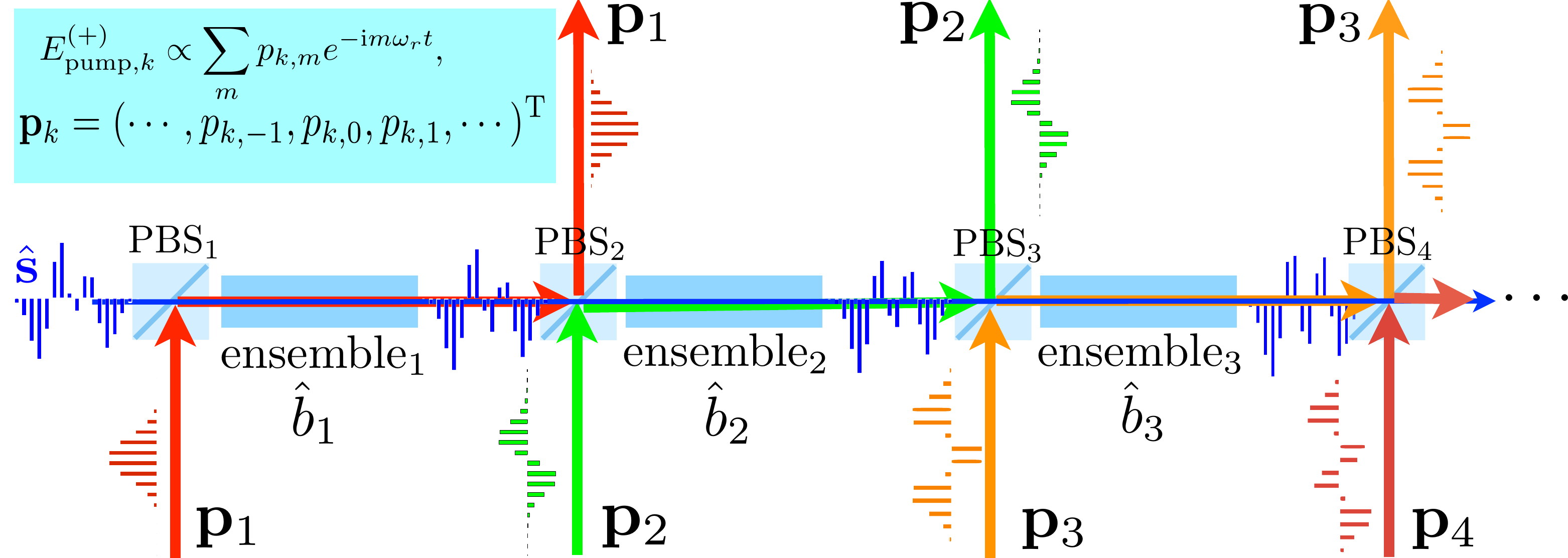}
\caption{\footnotesize (Color online) Quantum memories with different orthonormal pump field modes $\{{\bf p}_k\}_{k=1}^M$ concatenated to store and retrieve an $M$-intrinsic multimode light beam. 
}\label{figMemories}
\end{figure}
%===========================AtomicEnsemble===============================%

One of the main interests in the frequency combs is that they can be produced in highly multimode quantum states
that cannot be reduced to single modes. This is why we now consider the problem of storing and retrieving a multimode frequency comb, more precisely, as in the experimental situation \cite{Roslund2014,Pinel,Medeiros}, a quantum state consisting of $M$ uncorrelated squeezed states defined in orthonormal specific supermodes $\{{\pmb\psi}_m\}_{m=1}^{M}$. For this purpose, we will use $M$ atomic ensembles interacting in a row with the same signal comb, each one being pumped by a different pump field mode $\{{\bf p}_k\}_{k=1}^M$, which span the same sub-Hilbert space as the one by $\{{\pmb\psi}_k\}_{k=1}^M$,
\begin{equation}
\sum_{m=1}^M{\pmb\psi}_m{\pmb\psi}_m^{\text{H}}=\sum_{m=1}^M{\bf p}_m{\bf p}_m^{\text{H}}=\openone_{M},
\end{equation}
where $\openone_{M}$ is the identity operator acting on the sub-Hilbert space (see Fig. \ref{figMemories}). The signal field and the pump comb fields are assumed to be in orthogonal polarizations, so that they can be mixed and separated by polarizing beam splitters. Owing to the orthogonality between different pump modes and the far-off resonant Raman interaction, the memory processes are synchronous in each ensemble according to their own local times, therefore the operation times of the processes are identical, and the total input-output relation is a generalization of Eq. (\ref{in-out}), 
\begin{alignat}{2}
\underline{\hat{\bf s}}^{\text{out}}(\omega)=&-\mathcal{K}_{\omega}e^{\text{i}\omega T}\!\sum_{k=1}^M{\bf p}_k{\bf p}_k^{\text{H}}\underline{\hat{\bf s}}^{\text{in}}(\omega) &&\notag\\
=&-\mathcal{K}_{\omega}e^{\text{i}\omega T}\!\openone_M\underline{\hat{\bf s}}^{\text{in}}(\omega).&&\label{inoutens}
\end{alignat}
Note that it might be difficult to generate classical pump fields exactly matching the supermodes \{${\pmb\psi}_m$\}, by pulse-shaping techniques, for example. Instead, an advantage of our scheme is that any other complete sets of modes \{${\bf p}_m$\}, which are easier to prepare in practice, can be used to get the same output signal, as indicated by Eq. (\ref{inoutens}).

In experiments, the field quadratures of $m$th supermode are measured via operators $\hat{S}^{+}_{m}={\pmb\psi}_m^{\text{H}}\hat{\bf s}+\text{H.c.}, \hat{S}^{-}_{m}=-\text{i}{\pmb\psi}_m^{\text{H}}\hat{\bf s}+\text{H.c.}$, for example, with a multipixel homodyne detection method \cite{Pinel}. The covariance matrix of the field quadratures in the memory process is defined as
\begin{alignat}{2}
\mathcal{C}(\omega)&=\int_{\mathbb{R}}\frac{\text{d}\omega'}{2\pi}\langle 0|:\hat{\bf X}_{\omega}\hat{\bf X}_{\omega'}^{\text{T}}:|0\rangle+\mathbb{I}_{\text{2M}}, & &\\
\hat{\bf X}_{\omega} &= \int_{\mathbb{R}}\text{d}t e^{-\text{i}\omega t}\bigoplus_{m=1}^{M}\left(\begin{matrix}
  \hat{S}^{+}_{m}\\
  \hat{S}^{-}_{m}
\end{matrix}\right), &&
\end{alignat}
where the two colons stand for the normal ordering operation, and $\bigoplus$ represents the direct sum. Therefore, the covariance matrices $\mathcal{C}(\omega)$ of the write-in beam and the read-out beam yields
\begin{equation}
\mathbb{I}_{2\text{M}}-\mathcal{C}_{\text{out}}(\omega)=|\mathcal{K}_{\omega}|^2[\mathbb{I}_{2\text{M}}-\mathcal{C}_{\text{in}}(\omega)].\label{similarity}
\end{equation}
This relation constitutes the main result of this Rapid Communication: The two terms related to identity matrix $\mathbb{I}_{2\text{M}}$ in (\ref{similarity}) represents the contributions from vacuum inputs of atoms and light, as well as from the fluctuations of Langevin noises, while $|\mathcal{K}_{\omega}|^2\mathcal{C}_{\text{in}}(\omega)$ shows the amount of squeezing and anti-squeezing remaining at the read-out. The scaling factor $|\mathcal{K}_{\omega}|^2$ is insensitive to Fourier frequency $\omega$ if (1) $|\omega|\ll\gamma_s$, and (2) the optical depth $d\geq10$ when $|\omega|\leq\gamma_s$. In an experiment of the $ ^{171}$Yb$^+$ ions, $\Delta/2\pi=9$ THz, $\omega_r=80$ MHz, and $\gamma/2\pi=20$ MHz \cite{Hayes}. Let us take the peak Rabi frequency $|\Omega_p|/2\pi\simeq0.27$ THz of the picosecond pulses, and an atomic coherence lifetime of 0.1 s \cite{Radnaev}, which can also be at a minute scale for an atomic ensemble \cite{Olmschenk2007,Dudin2013}, and can be much longer \cite{Maurer,Saeedi}, even of 6 h \cite{Zhong} with solid-state materials in experiments. The off-resonant Raman memory model is still valid under these conditions \cite{Golubeva2012}. Then, the induced decay rate $\gamma_s/2\pi\simeq10$ kHz, and at most 0.2\% errors of the efficiencies are introduced by the finite interaction time if one takes $d=4$ and $T=1$ ms, which enables one to store about $10^5$ pulses, or $T=0.1$ ms when $d\geq14$. In general, when the pulse duration $T\geq10\times2\pi\gamma_s^{-1}$ but still smaller than the atomic coherence time, all the spectral components of the slowly varying envelope of the pulse lay within $|\omega|\leq0.1\gamma_s$, and the scaling factor $|\mathcal{K}_{\omega}|^2$ can be replaced by its value at $\omega=0$. This is what we will do in the remaining part of the Rapid Communication. 

 The positive eigenvalues of $\mathbb{I}_{2\text{M}}-\mathcal{C}$ reveal the squeezing in a Gaussian state \cite{Arvind}. Under mode basis $\{{\pmb \psi}_k\}_{k=1}^M$, the covariance matrix $\mathcal{C}$ turns out to be $\bigoplus_{k=1}^M\mathcal{C}^{{\pmb \psi}_k}$, which is diagonal with $\mathcal{C}^{{\pmb \psi}_k}$ being the block on mode ${\pmb\psi}_k$. Therefore, the relation of the amount of the squeezing between the read-out and write-in state is determined by $\mathcal{K}_{0}^2$ according to Eq. (\ref{similarity}), and leads to the memory efficiency \cite{Kozhekin2000} of any mode ${\pmb\psi}_m$,
\begin{equation}
\eta=\mathcal{K}_0^2=(1-e^{-d})^2,
\end{equation}
close to 1 even for moderate values of the on-resonance optical depth. The amount of squeezing $\zeta$ in the retrieved $m$-th supermode is 
\begin{equation}
\zeta^{(m)}_{\text{out}}=1-\eta(1-\zeta_{\text{in}}^{(m)}).
\end{equation}

%-----------------------------------------------------------------------------------------
 %-----------------------------------------------------------------------------------------
  \begin{figure}[htbp] 
   \centering  
   \includegraphics[scale=1.15]{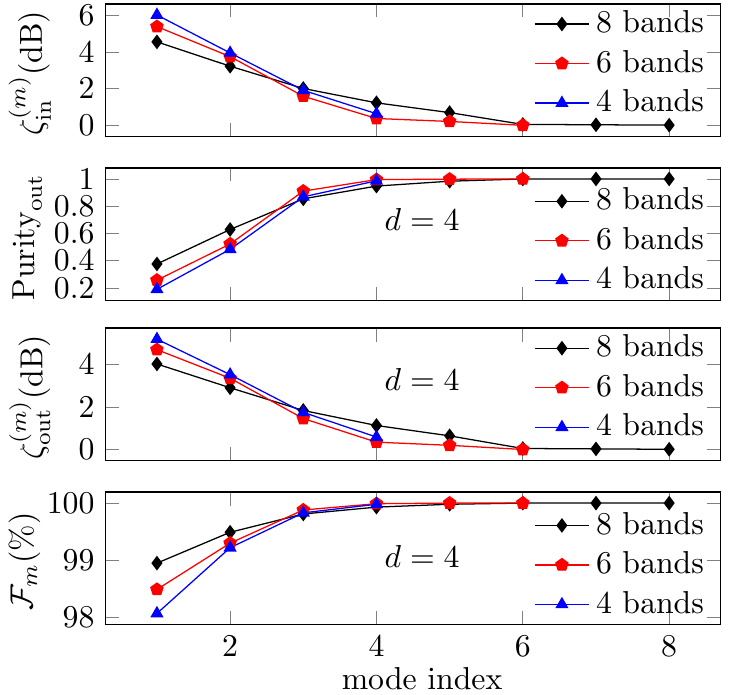}
   \caption{\footnotesize(Color online) Quantum memory for the squeezed frequency comb. From top to bottom: Input squeezing in supermodes (from Ref. \cite{Medeiros}), purity of the read-out light, squeezing of the read-out light, fidelity of the memory. One notes that a supermode with less input squeezing is better retrieved at the output.}
   \label{squeezing:inout}
 \end{figure}
%-----------------------------------------------------------------------------------------
%-----------------------------------------------------------------------------------------
This expression shows that the squeezing is well preserved in the write/read process when $\eta$ is close to 1. The product of field quadrature variances of the $m$th supermode, $\text{det} \mathcal{C}_{\text{out}}^{{\pmb\psi}_m}=1+\eta(1-\eta)[\text{Tr}\sqrt{\mathcal{C}^{{\pmb\psi}_m}_{\text{in}}}\text{diag}(1,-1)]^2$, attains the Heisenberg limit 1 only at large optical depth, meaning that the purity of the read-out state in mode ${\pmb\psi}_m$ is slightly degraded, in the same proportion as the loss of squeezing. Finally, the fidelity $\mathcal{F}=\big(\text{Tr}\sqrt{\rho^{1/2}_{\text{in}}\rho_{\text{out}}\rho^{1/2}_{\text{in}}}\big)^2$ for squeezed vacuum states on mode ${\pmb\psi}_m$ is
\begin{equation}
\mathcal{F}_m=2[4+(1-\eta^2)\text{Tr}(\mathcal{C}_{\text{in}}^{{\pmb\psi}_m}-\mathbb{I}_2)]^{-1/2}. \label{Fidelity}
\end{equation}
It depends not only on the optical depth but also on the variances of the initial signal comb field. Ideally, when the optical depth is large, or the initial state is a multimode coherent state, the fidelity is close to 1. Note that the most squeezed supermode in the input state has the worst fidelity at the output according to Eq. (\ref{Fidelity}); the overall fidelity $\mathcal{F}=\prod_m\mathcal{F}_m$ can be low at a given optical depth when (1) the fidelity of the most squeezed supermode is poor, and (2) the comb state is quite complex, and therefore should be characterized by a large number of supermodes $(M\gg1)$. However, the overall fidelity is not a precise quantity to describe the memory process, so it is more proper to use a vector ($\mathcal{F}_1,\cdots,\mathcal{F}_M$), because in general some supermodes are more squeezed, and they should have more weight. To faithfully store and retrieve an intrinsic high-dimensional multimode comb state, one has to implement every ensemble at a large optical depth.

To theoretically illustrate the protocol we consider a set of memories working with an intrinsic $M$-mode pure squeezed vacuum state that is generated by a synchronously pumped optical parametric oscillator device below but close to threshold \cite{SPOPO}, and the squeezing data in each supermodes of different states (see the top graph in Fig. \ref{squeezing:inout}). The purity, squeezing, and fidelity of each supermode of different retrieval states at optical depth 4, are shown in the remaining three graphs in Fig. \ref{squeezing:inout}. To have better purity, squeezing and higher fidelity of the retrieved light, one can choose ensembles with a larger optical depth, for example, $d\simeq20$ \cite{Kozhekin2000}, and $d\simeq1800$ \cite{Reim2010}.  

Any pure multimode Gaussian state can be reduced by a mode transformation as a tensor product of squeezed vacuum states on different orthogonal modes \cite{Arvind,Braunstein}. Therefore, the previous scheme for storing several uncorrelated squeezed vacuum states in different modes is also a scheme for storing multimode entanglement in another mode basis. We thus have the ability to store multimode entanglement in comb states, including the continuous variable cluster states, by using this scheme.

In this Rapid Communication, we have shown a possible way to store the multipartite entanglement in the quantum optical frequency comb state, by using a set of memories driven with different classical comb light beams. We also showed that the purity, fidelity, and efficiency are very close to 1 at a large, but experimentally feasible, optical depth, and that the crucial element to store and retrieve a multimode state is the optical depth. We have also shown that it is possible to manipulate a large set of entangled modes in signal light by using a series of independent atomic ensembles.

We acknowledge fruitful discussions with J. Mueller, A. S{\o}rensen, A. Grodecka-Grad, and J. Laurat, and the financial support of the Future and Emerging Technologies (FET) programme within the Seventh Framework Programme for Research of the European Commission, under the FET-Open Grant Agreement HIDEAS, No. FP7-ICT-221906, the individual Marie Curie (IEF) project AAPLQIC (Grant No. 330004), and the European Research Council starting grant Frecquam. C.F. acknowledges support as a member of the Institut Universitaire de France, Z.Z. acknowledges the China Scholarship Council for support.

\end{document}